\documentclass[showpacs,preprintnumbers,twocolumn,amsmath,amssymb]{revtex4}%
\usepackage{epsfig}
\usepackage{subfigure}
\usepackage{graphicx}
\usepackage{dcolumn}
\usepackage{stmaryrd}
\usepackage{mathrsfs}
\usepackage{pifont}
\usepackage{amsthm}
\usepackage{bm}
\usepackage{latexsym}
\usepackage{amsmath}
\usepackage{amsfonts}
\usepackage{amssymb}%
\providecommand{\U}[1]{\protect\rule{.1in}{.1in}}

\newcommand{\beq}{\begin{equation}}
\newcommand{\eeq}{\end{equation}}
\newcommand{\beqa}{\begin{eqnarray}}
\newcommand{\eeqa}{\end{eqnarray}}

\begin{document}
\title{Dynamics of interacting qubits coupled to a common bath: Non-Markovian quantum
state diffusion approach}
\author{Xinyu Zhao%
\footnote{Email:xzhao1@stevens.edu}%
, Jun Jing, Brittany Corn and Ting Yu%
\footnote{Email:Ting.Yu@stevens.edu}%
}
\affiliation{Center for Controlled Quantum Systems and the Department of Physics and
Engineering Physics, Stevens Institute of Technology, Hoboken, New Jersey
07030, USA}

\begin{abstract}
Non-Markovian dynamics is studied for two interacting quibts strongly coupled
to a dissipative bosonic environment. For the first time, we have derived the
non-Markovian quantum state diffusion (QSD) equation for the coupled two-qubit
system without any approximations, and in particular, without the Markov
approximation. As an application and illustration of our derived time-local
QSD equation, we investigate the temporal behavior of quantum coherence
dynamics. In particular, we find a strongly non-Markovian regime where
entanglement generation is significantly modulated by the environmental
memory. Additionally, we studied the residual entanglement in the steady state
by analyzing the steady state solution of the QSD equation. Finally, we have
discussed an approximate QSD equation.

\end{abstract}

\maketitle


\section{Introduction}

The theory of quantum open systems has attracted growing attention because the
framework of system plus environment has become a fundamental paradigm in
quantum dissipative system, quantum optics and quantum information science
\cite{Openbook1,Openbook2}. In reality, no quantum systems can be completely
isolated from their environment. The mutual interaction between the system and
environment is responsible for many important physical processes such as
dissipation, fluctuation, as well as decoherence and disentanglement. Quantum
open system is also an essential part of quantum information processing where
the quantum noise and control are major issues in realizing quantum computing,
quantum teleportation and quantum cryptography \cite{Nielsen}.

While the theory of open systems in the Markov regimes can be extensively
treated with the standard Lindblad master equations \cite{Openbook1} or the
corresponding Markov quantum trajectories
\cite{Carmicheal,Dalibardetal,Gisin-Percival,Knight}, the microscopic
non-Markovian theory has not been well developed for the systems consisting of
interacting spins, which are of experimental interest \cite{Xu}. The
difficulty arising from the strong coupling between the system and its
environment is that the non-Markovian effects due to action and back reaction
between the system and the environment must be taken into account.
Undoubtedly, some memory related properties such as quantum entanglement,
inhibited spontaneous emission, quantum transport, quantum tunneling
\cite{Bellomo,Yu-Eberly10,Vats,Huetalnew1,zhangweimin,Yangroup1,Yangroup2,ShaoJS,Tan}
which are of importance will be lost when simply taking the Markov limit.
Among all of the techniques attempt to investigate the non-Markovian
evolution, the quantum state diffusion (QSD) equation initially proposed by
Di\'{o}si, Gisin, and Strunz \cite{QSD} provides a powerful tool to deal with
the quantum open systems coupled to a boson bath \cite{QSD2,Strunz,QBM}.

The recent developments of the non-Markovian dynamics are concerned with a
theoretical derivation of the QSD equation directly from an underlying
microscopic model irrespective of environmental memory time and coupling
strength \cite{QSD}. The non-Markovian QSD equation is equivalent to the
master equation formalism, but offers numerical advantages over the
corresponding master equation if it exists. Typical to non-Markovian evolution
is that the dynamics is explicitly dependent on the past history manifested in
the integrals over the time in the dynamic equation
\cite{Nakajima,Zwanzig,Haake}. In the case of non-Markovian QSD, under certain
conditions, the nonlocal integral can be cast into a time-local form
\cite{QSD,QBM,Jing-Yu2010,Corn}. Then, the time-local QSD equation can be
efficiently implemented as an analytical and numerical tools to describe the
non-Markovian effects.

The primary purpose of this paper is to study the non-Markovian quantum
dynamics of interacting qubits. We derive a closed exact time-local QSD
equation for two interacting qubits coupled to a bosonic environment by
replacing the functional derivative in the QSD equation with a time-dependent
operator called $O$ operator. We have established the explicit equations of
motion for the coefficient functions contained in the $O$ operator. The second
purpose of this paper is to investigate the non-Markovian dynamics of qubit
systems by using the derived exact QSD equation. Since the exact QSD equation
is valid for an arbitrary structured boson bath and the strong coupling
regime, we are thus able to address the important issue of entanglement
generation via environment memory in the full non-Markovian regime. It is
shown that several interesting features may arise from this coupled qubit
system when we examine the dynamics in a non-Markovian regime. With this
coupled qubit model, we have studied the approximation based on the functional
expansion of the $O$ operator, and reveal some interesting features on
non-Markovian perturbation theory.

The organization of this paper is as follows. In Sec. II, we introduce an
interacting qubit model and develop the theory of the exact time-local quantum
state diffusion equation. In Sec. III, based on the exact QSD equation
derived, we consider memory-assisted entanglement to illustrate the
implications of non-Markovian dynamics on entanglement generation. As an
example, we show that entanglement generation may be significantly modified by
the environment memory time. With the quantum trajectories generated by the
exact QSD equation, we also analyze the steady state solution for the single
trajectories recovering the steady state entanglement described by the density
matrix. We conclude in Sec. IV.

\section{The Model and Time-Local Quantum State Diffusion Equation}

The model consists of a pair of coupled two-level atoms (or spins) interacting
with a common bath. The total Hamiltonian can be written as (setting $\hbar
=1$),
\begin{equation}
H_{\mathrm{tot}}=H_{\mathrm{sys}}+H_{\mathrm{env}}+H_{\mathrm{int}},
\label{Hamiltonian}%
\end{equation}
where
\begin{align}
H_{\mathrm{sys}}  &  =\omega_{A}\sigma_{z}^{A}+\omega_{B}\sigma_{z}^{B}%
+J_{xy}(\sigma_{+}^{A}\sigma_{-}^{B}+\sigma_{-}^{A}\sigma_{+}^{B})+J_{z}%
\sigma_{z}^{A}\sigma_{z}^{B},\nonumber\\
H_{\mathrm{env}}  &  =\sum_{j}\omega_{j}b_{j}^{\dagger}b_{j},\nonumber\\
H_{\mathrm{int}}  &  =\sum_{j}(g_{j}b_{j}^{\dagger}L+g_{j}^{\ast}%
b_{j}L^{\dagger}),
\end{align}
where $L=\kappa_{A}\sigma_{-}^{A}+\kappa_{B}\sigma_{-}^{B}$ is the system
Lindblad operator coupled to the environment. Note that $\kappa_{A},\kappa
_{B}$ are constants describing different coupling strengths for two spins.

This type of interaction between qubits has been studied widely in the
Heisenberg spin chain models \cite{spin chain-start,spin chain1,spin
chain-end} and proved to be a useful model in quantum information processing
based on spin chain. Moreover, some cavity-QED systems may be described by a
similar type of Hamiltonian in simulating the Heisenberg spin chain
\cite{cav-spin}. Notably, our model in the framework of quantum open system is
a very general case, including many other physically interesting models as
special cases of this Hamiltonian. For example, when $J_{z}=0$, this model
will reduce to the example of the Heisenberg $XX$ type interaction. When
$J_{xy}=J_{z}=0$, two qubits will be treated as non-interacting spin. When
$\omega_{B}=J_{xy}=J_{z}=\kappa_{B}=0,$ this model simply reduces to the one
qubit dissipative model \cite{QSD2}.

The stochastic Schr\"{o}dinger equation for this interacting spin model may be
formally written as \cite{QSD,QSD2,QBM},
\begin{equation}
\frac{\partial}{\partial t}\psi_{t}=-iH_{\mathrm{sys}}\psi_{t}+Lz_{t}^{\ast
}\psi_{t}-L^{\dagger}%
{\displaystyle\int\nolimits_{0}^{t}}
ds\alpha(t,s)\frac{\delta\psi_{t}}{\delta z_{s}^{\ast}}, \label{SSE}%
\end{equation}
where $\alpha(t,s)=\sum_{j}|g_{j}|^{2}e^{-i\omega_{j}(t-s)}$ is the bath
correlation function and $z_{t}^{\ast}=-i\sum_{j}g_{j}^{\ast}z_{j}^{\ast
}e^{i\omega_{j}t}$ is a complex Gaussian process satisfying $M[z_{t}%
]=M[z_{t}z_{s}]=0$, and $M[z_{t}^{\ast}z_{s}]=\alpha(t,s)$. Here $M[\cdot]$
denotes the ensemble average over the classical noise $z_{t}$. The quantum
trajectory $\psi_{t}(z^{\ast})$ recovers the density operator of the system in
the ensemble average: $\rho_{t}=M[|\psi_{t}(z^{\ast})\rangle\langle\psi
_{t}(z^{\ast})|]=\int\frac{dz^{2}}{\pi}e^{-|z|^{2}}|\psi_{t}(z^{\ast}%
)\rangle\langle\psi_{t}(z^{\ast})|$.

The above exact equation contains a time-nonlocal term, it renders practical
application impossible without evoking certain approximation. The existence of
the time-local equation depends on if one can replace the functional
derivative in the integral in Eq.~(\ref{SSE}) with a time-dependent (even
noise-dependent) operator $O(t,s,z^{\ast})$:
\begin{equation}
\frac{\delta\psi_{t}(z^{\ast})}{\delta z_{s}^{\ast}}=O(t,s,z^{\ast})\psi
_{t}(z^{\ast}).
\end{equation}

Using the \textquotedblleft consistency condition" \cite{QSD}
\begin{equation}
\frac{\delta}{\delta z_{s}^{\ast}}\frac{\partial\psi_{t}}{\partial t}%
=\frac{\partial}{\partial t}\frac{\delta\psi_{t}}{\delta z_{s}^{\ast}},
\end{equation}
one can obtain a formal evolution equation for the operator $O(t,s,z^{\ast}%
)$:
\begin{equation}
\frac{\partial}{\partial t}O=[-iH_{\mathrm{sys}}+Lz_{t}^{\ast}-L^{\dagger}%
\bar{O},O]-L^{\dagger}\frac{\delta}{\delta z_{s}^{\ast}}\bar{O},
\label{consistency condition}%
\end{equation}
where $\bar{O}(t,z_{{}}^{\ast})\equiv%
{\displaystyle\int_{0}^{t}}
ds\alpha(t-s)O(t,s,z^{\ast}).$ This equation of motion for the $O$ operator
has to be solved with the initial condition,
\begin{equation}
O(t,s=t,z_{{}}^{\ast})=L. \label{ini. cond. for O}%
\end{equation}

The explicit $O$ operator in Eq.~(\ref{consistency condition}) has been
determined for several interesting models \cite{QSD,QBM,Jing-Yu2010}. However,
the construction of the $O$ operator for the coupled two spins has not been
found before. It is generally very difficult to find the exact $O$ operator.
As the major result of this paper, we have found the exact $O$ operator for
this coupled qubit model. To begin with, we note that the $O$ operator may be
expaned in the following way,
\begin{align}
O(t,s,z^{\ast})  &  =f_{1}(t,s)O_{1}+f_{2}(t,s)O_{2}+f_{3}(t,s)O_{3}%
\nonumber\\
&  +f_{4}(t,s)O_{4}+i\int_{0}^{t}ds^{\prime}f_{5}(t,s,s^{\prime})z_{s^{\prime
}}^{\ast}O_{5}, \label{Ansztz of O}%
\end{align}
where the basis operators are given by
\begin{align}
&  O_{1}=\sigma_{-}^{A},\,\,O_{2}=\sigma_{-}^{B},\,\,O_{3}=\sigma_{z}%
^{A}\sigma_{-}^{B},\nonumber\\
&  O_{4}=\sigma_{z}^{B}\sigma_{-}^{A},\,\,O_{5}=2\sigma_{-}^{A}\sigma_{-}^{B},
\end{align}
and $f_{j}$ $(j=1,2,3,4,5)$ are some time-dependent coefficients. Substituting
Eq. (\ref{Ansztz of O}) into Eq. (\ref{consistency condition}), we obtain the
partial differential equations governing the coefficients of the $O$ operator.%
\begin{widetext}%
%

\begin{align}
\frac{\partial}{\partial t}f_{1}(t,s)  &  =+2i\omega_{A}f_{1}+\kappa_{A}%
F_{1}f_{1}+\kappa_{B}F_{3}f_{1}-iJ_{xy}f_{3}-\kappa_{B}F_{1}f_{3}+\kappa
_{B}F_{4}f_{3}+2iJ_{z}f_{4}+\kappa_{A}F_{4}f_{4}+\kappa_{B}F_{3}f_{4}%
-i\kappa_{B}F_{5},\nonumber\label{diff. of O}\\
\frac{\partial}{\partial t}f_{2}(t,s)  &  =+2i\omega_{B}f_{2}+\kappa_{A}%
F_{4}f_{2}+\kappa_{B}F_{2}f_{2}+2iJ_{z}f_{3}+\kappa_{A}F_{4}f_{3}+\kappa
_{B}F_{3}f_{3}-iJ_{xy}f_{4}-\kappa_{A}F_{2}f_{4}+\kappa_{A}F_{3}f_{4}%
-i\kappa_{A}F_{5},\nonumber\\
\frac{\partial}{\partial t}f_{3}(t,s)  &  =-iJ_{xy}f_{1}-\kappa_{A}F_{2}%
f_{1}+\kappa_{A}F_{3}f_{1}+2iJ_{z}f_{2}+\kappa_{A}F_{4}f_{2}+\kappa_{B}%
F_{3}f_{2}+2i\omega_{B}f_{3}+\kappa_{A}F_{4}f_{3}+\kappa_{B}F_{2}f_{3}%
-i\kappa_{A}F_{5},\nonumber\\
\frac{\partial}{\partial t}f_{4}(t,s)  &  =+2iJ_{z}f_{1}+\kappa_{A}F_{4}%
f_{1}+\kappa_{B}F_{3}f_{1}-iJ_{xy}f_{2}-\kappa_{B}F_{1}f_{2}+\kappa_{B}%
F_{4}f_{2}+2i\omega_{A}f_{4}+\kappa_{A}F_{1}f_{4}+\kappa_{B}F_{3}f_{4}%
-i\kappa_{B}F_{5},\nonumber\\
\frac{\partial}{\partial t}f_{5}(t,s,s^{\prime})  &  =+\kappa_{A}F_{5}%
f_{1}+\kappa_{B}F_{5}f_{2}-\kappa_{B}F_{5}f_{3}-\kappa_{A}F_{5}f_{4}%
+2i\omega_{A}f_{5}+2i\omega_{B}f_{5}+\kappa_{A}F_{1}f_{5}+\kappa_{A}F_{4}%
f_{5}+\kappa_{B}F_{2}f_{5}+\kappa_{B}F_{3}f_{5},
\end{align}%
\end{widetext}%
where $F_{j}(t)=\int_{0}^{t}ds\alpha(t,s)f_{j}(t,s)$ $(j=1,2,3,4)$ and
$F_{5}(t,s^{\prime})=\int_{0}^{t}ds\alpha(t,s)f_{5}(t,s,s^{\prime})$, with the
initial conditions for Eq. (\ref{diff. of O}):
\begin{align}
f_{1}(t,s  &  =t)=\kappa_{A},\nonumber\\
f_{2}(t,s  &  =t)=\kappa_{B},\nonumber\\
f_{3}(t,s  &  =t)=0,\nonumber\\
f_{4}(t,s  &  =t)=0,\nonumber\\
f_{5}(t,s  &  =t,s^{\prime})=0,\nonumber\\
f_{5}(t,s,s^{\prime}  &  =t)=-i[\kappa_{A}f_{3}(t,s)+\kappa_{B}f_{4}(t,s)].
\end{align}
The time evolution for this model can be solved numerically with the
coefficients determined by Eq. (\ref{diff. of O}).

With the explicit $O$ operator (\ref{Ansztz of O}), the linear QSD equation
can be written compactly as a time-local equation:
\begin{equation}
\frac{\partial}{\partial t}\psi_{t}=-iH_{\mathrm{sys}}\psi_{t}+Lz_{t}^{\ast
}\psi_{t}-L^{\dagger}\bar{O}(t,z^{\ast})\psi_{t}, \label{tlSSE}%
\end{equation}
where
\begin{equation}
\bar{O}(t,z^{\ast})=\sum_{j=1}^{4}F_{j}(t)O_{j}+i\int_{0}^{t}ds^{\prime}%
F_{5}(t,s^{\prime})z_{s^{\prime}}^{\ast}O_{5}.
\end{equation}

Although the time evolution is, in principle, governed by the linear QSD
equation (\ref{SSE}), it is much more efficient to use the nonlinear version
in numerical simulations since it preserves the norm of the state vector. The
nonlinear QSD equation can be written as \cite{QSD},
\begin{align}
\frac{\partial}{\partial t}\widetilde{\psi}_{t}  &  =-iH_{\mathrm{sys}%
}\widetilde{\psi}_{t}+(L-\left\langle L\right\rangle _{t})\widetilde{z}%
_{t}^{\ast}\widetilde{\psi}_{t}\nonumber\\
&  -[(L^{\dagger}-\left\langle L^{\dagger}\right\rangle _{t})\bar
{O}-\left\langle (L^{\dagger}-\left\langle L^{\dagger}\right\rangle _{t}%
)\bar{O}\right\rangle ]\widetilde{\psi}_{t}, \label{NNMQSD}%
\end{align}
where $\widetilde{\psi}_{t}=\frac{\psi_{t}}{\left\Vert \psi_{t}\right\Vert }$
is the normalized state vector , $\left\langle L\right\rangle _{t}%
=\left\langle \widetilde{\psi}_{t}\right\vert L\left\vert \widetilde{\psi}%
_{t}\right\rangle $ denotes the mean value of the operator and $\widetilde
{z}_{t}^{\ast}=z_{t}^{\ast}+\int_{0}^{t}ds\alpha^{\ast}(t,s)\left\langle
L^{\dagger}\right\rangle _{s}$ is the shifted noise.

Let us emphasize that our results are valid for an arbitrary correlation
functions. For simplicity, in this paper, we model the noise as an
Ornstein-Uhlenbeck process for which the correlation function is
$\alpha(t,s)=\frac{\gamma}{2}e^{-\gamma\left\vert t-s\right\vert }$. This
correlation function is very useful in demonstrating the non-Markovian effects
as well as the Markov case by varying the parameter $\gamma$, which controls
the memory effect of the reservoir. When $\gamma\rightarrow\infty$,
$\alpha(t,s)\rightarrow\delta(t-s)$ it corresponds to the Markov case.
Oppositely, when $\gamma$\ is small, the system is in a non-Markovian regime.

\section{Numerical Results and Discussions}

Having developed the time-local non-Markovian QSD equation for the two
interacting qubits, this section is concerned with the non-Markovian dynamics
of entanglement involving entanglement generation and evolution
\cite{Yu-Eberly02,ET6,ET7,ET8,ET9,Ficek09,Strunz2009,CPSun2007,CPSun2008}.
Throughout the paper, the degree of entanglement is measured by "concurrence"
\cite{concurrence}.

\subsection{Memory-assisted entanglement generation}

\begin{figure}[ptb]
\begin{center}
\includegraphics[
trim=0.000000in 0.000000in -0.836388in 0.000000in,
height=2.2087in,
width=3.0113in
]{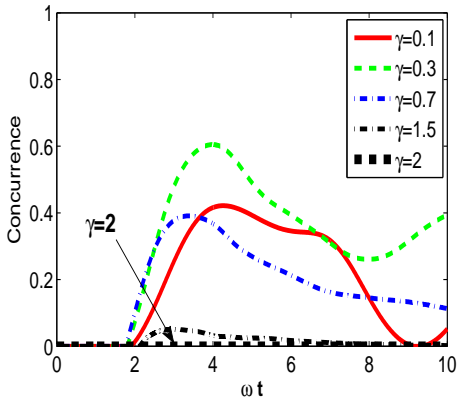}
\end{center}
\caption{Entanglement generation for different memory times of the noise. A
particular initial state is chosen as $\psi_{0}=\left\vert 11\right\rangle .$
The curves for different $\gamma$ are marked in the graph. The other
parameters are $\omega_{A}=\omega_{B}=0.5\omega,$ $\kappa_{A}=\kappa_{B}=1$,
$J_{xy}=0.5.$ }%
\label{noise1}%
\end{figure}

\begin{figure}[ptb]
\begin{center}
\includegraphics[
trim=0.000000in 0.000000in 0.000000in -0.482355in,
height=2.3091in,
width=3.0113in
]{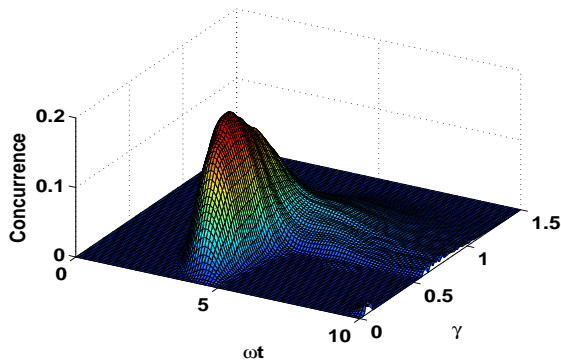}
\end{center}
\caption{Entanglement generation vs $\gamma$ and $t$. A particular initial
state is chosen as $\psi_{0}=\left\vert 11\right\rangle .$ The other
parameters are $\omega_{A}=\omega_{B}=0.5\omega,$ $\kappa_{1}=\kappa_{B}=1$,
$J_{xy}=J_{z}=0.$ }%
\label{noise2}%
\end{figure}

With our exact non-Markovian QSD equation (\ref{NNMQSD}), we are able to
consider the important non-Markovian feature of entanglement generation. For
this purpose, we consider the separable state $\psi_{0}=\left\vert
11\right\rangle $ to show how the environmental memory affects the
entanglement generation. It has been known that the entanglement may be
generated for a qubit system through coupling qubits to a common bath.
However, it is not clear how the environmental correlation times affects
entanglement generation. In Fig.~\ref{noise1}, we plot the entanglement
evolution for different parameters of $\gamma,$ where $1/\gamma$ gives the
memory time of the environment. With this initial state, we can see that when
$\gamma\geq2$ (close to the Markov limit for this model), there will be no
visible entanglement generated for all times. In fact, it is easy to show that
in the Markov limit no entanglement will be generated (i.e. $\gamma
\rightarrow\infty$) for this special initial state. In addition, it is seen
that when the parameter is too small, the entanglement generation becomes less
significant shown in our numerical simulations. This shows that the
non-Markovian property affects generating entanglement in a subtle way.

We can show that entanglement generation for this special initial state is due
to the environmental memory rather than the spin-spin coupling. Interestingly
enough, we find that the largest amount of entanglement may be generated if
the correlation time of environment is neither too long ($\gamma$ is small)
nor too short ($\gamma$ is large). In Fig. \ref{noise2}, we plot the case
without coupling between the two spins. The results confirm that entanglement
for the two non-interacting qubits is purely generated by environment memory.
Our results here have clearly shown that non-Markovian noise can be very
useful rather than harmful to the entanglement if the memory times are
properly chosen.


\subsection{Steady state entanglement}

\begin{figure}[ptb]
\begin{center}
\includegraphics[
trim=0.000000in 0.000000in 0.000000in -0.284798in,
height=1.8066in,
width=3.2119in
]{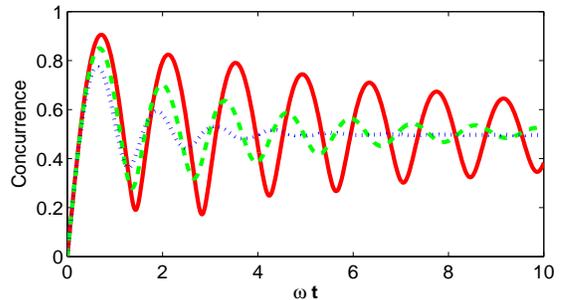}
\end{center}
\caption{Entanglement generation$.$ The red (solid), green (dashed) and blue
(short-dashed) lines represents $\gamma=0.5,$ $\gamma=1,$ $\gamma=2,$
respectively. The initial state is $\psi_{0}=\left\vert 10\right\rangle .$ The
other parameters are $\omega_{A}=\omega_{B}=0.5\omega,$ $J_{xy}=0.5,$
$J_{z}=0.1,$ $\kappa_{A}=\kappa_{B}=1.$}%
\label{entgen}%
\end{figure}Quantum trajectories generated by the non-Markovian QSD equation
(\ref{SSE}) have many interesting properties. For instance, we can use those
quantum trajectories to study the steady state entanglement. In Fig.
\ref{entgen}, we plot the entanglement dynamics from a separable initial state
$\psi_{0}=\left\vert 10\right\rangle $. The numerical results show that the
initial pure state evolves into a steady entangled state. In fact, with the
help of non-Markovian trajectories, we can show that all the trajectories will
localize to the eigenstates of the Lindblad operator $L$ and the residual
entanglement of the steady state is not zero. Now, let us examine the large
$t$ behavior of the single trajectories. First, we choose the following four
states%
\begin{align}
\left\vert \psi_{1}\right\rangle  &  =\left\vert 11\right\rangle ,\nonumber\\
\left\vert \psi_{2}\right\rangle  &  =\frac{1}{\sqrt{2}}(\left\vert
10\right\rangle +\left\vert 01\right\rangle ),\nonumber\\
\left\vert \psi_{3}\right\rangle  &  =\frac{1}{\sqrt{2}}(\left\vert
10\right\rangle -\left\vert 01\right\rangle ),\nonumber\\
\left\vert \psi_{4}\right\rangle  &  =\left\vert 00\right\rangle .
\end{align}
which form a complete basis of the Hilbert space. Hence, an arbitrary state
can always be written as
\begin{equation}
\left\vert \psi\right\rangle =c_{1}\left\vert \psi_{1}\right\rangle
+c_{2}\left\vert \psi_{2}\right\rangle +c_{3}\left\vert \psi_{3}\right\rangle
+c_{4}\left\vert \psi_{4}\right\rangle , \label{expand psi}%
\end{equation}
where $|c_{1}|^{2}+|c_{2}|^{2}+|c_{3}|^{2}+|c_{4}|^{2}=1.$

Substituting Eq. (\ref{expand psi}) into the QSD Eq. (\ref{tlSSE}), we get the
following stochastic differential equations for the coefficients:
\begin{align}
\frac{\partial}{\partial t}c_{1}(t)  &  =i(2\omega+J_{z})c_{1}-2F_{1}%
(t)c_{1}-2F_{3}(t)c_{1},\label{c1}\\
\frac{\partial}{\partial t}c_{2}(t)  &  =i(J_{xy}-J_{z})c_{2}+\sqrt{2}%
z_{t}^{\ast}c_{1}-2F_{1}(t)c_{2}\label{c2}\\
&  +2F_{3}(t)c_{2}-2\sqrt{2}\tilde{F}_{5}(t,z^{\ast})c_{1},\nonumber\\
\frac{\partial}{\partial t}c_{3}(t)  &  =-i(J_{xy}+J_{z})c_{3},\label{c3}\\
\frac{\partial}{\partial t}c_{4}(t)  &  =i(J_{z}-2\omega)c_{4}-\sqrt{2}%
z_{t}^{\ast}c_{2}. \label{c4}%
\end{align}
where $\tilde{F}_{5}(t,z^{\ast})=i\int_{0}^{t}ds^{\prime}F_{5}(t,s^{\prime
})z_{s^{\prime}}^{\ast}$. Here, we assume that the two qubits are identical
and symmetric ($\omega_{A}=\omega_{B},\kappa_{A}=\kappa_{B}$) for simplicity.
The more general case can be dealt with in a similar way. For the initial
state $\psi_{0}=|10\rangle$, the initial conditions for this set of equations
are
\begin{align}
c_{1}(0)  &  =0,\\
c_{2}(0)  &  =\frac{1}{\sqrt{2}},\\
c_{3}(0)  &  =\frac{1}{\sqrt{2}},\\
c_{4}(0)  &  =0.
\end{align}
For the above initial conditions, $c_{1}(t)\equiv0$ at any times, and the
steady state solution for $c_{2}(t)$ is simply given by
\begin{equation}
c_{2}(\infty)=0.
\end{equation}
Thus, the large $t$ state for a single trajectory will not contain $\left\vert
\psi_{1}\right\rangle $ and $\left\vert \psi_{2}\right\rangle $ components,
i.e.%
\begin{equation}
\left\vert \psi_{\mathrm{final}}\right\rangle =c_{3}\left\vert \psi
_{3}\right\rangle +c_{4}\left\vert \psi_{4}\right\rangle . \label{psi_f}%
\end{equation}
Now, it is clear that all trajectories will always evolve into a sub-space of
the Hilbert space spanned by $\left\vert \psi_{3}\right\rangle $ and
$\left\vert \psi_{4}\right\rangle $, which are the eigenstates of the Lindblad
operator $L=\sigma_{-}^{A}+\sigma_{-}^{B}$. Moreover, we can easily derive the
equations of the motion for the ensemble means of the cross terms. For
example, $M[c_{3}^{\ast}c_{4}]$ satisfies the following equation:
\[
\frac{d}{dt}M[c_{3}^{\ast}c_{4}]=(iJ_{xy}+2iJ_{z}-2i\omega)M[c_{3}^{\ast}%
c_{4}]-\sqrt{2}M[z_{t}^{\ast}c_{3}^{\ast}c_{2}].
\]
Since $c_{2}(\infty)=0$ in the long-time limit, we have
\begin{equation}
M[c_{3}^{\ast}c_{4}]\rightarrow0,\,\,M[c_{3}^{\ast}c_{4}]\rightarrow0.
\end{equation}
where $t$ is large. Therefore, the density matrix recovered by $\rho
_{t}=M[|\psi_{t}(z^{\ast})\rangle\langle\psi_{t}(z^{\ast})|]$ will approach to
the steady state density matrix in long-time limit:%
\begin{equation}
\rho_{s}\approx\left[
\begin{array}
[c]{cccc}%
0 & 0 & 0 & 0\\
0 & 0 & 0 & 0\\
0 & 0 & M[c_{3}^{\ast}c_{3}]|_{t\rightarrow\infty} & 0\\
0 & 0 & 0 & M[c_{4}^{\ast}c_{4}]|_{t\rightarrow\infty}%
\end{array}
\right]  ,
\end{equation}
in the basis $\{\left\vert \psi_{1}\right\rangle ,$ $\left\vert \psi
_{2}\right\rangle ,$ $\left\vert \psi_{3}\right\rangle ,$ $\left\vert \psi
_{4}\right\rangle \}$. The steady state density matrix in the basis
$\{\left\vert 11\right\rangle ,$ $\left\vert 10\right\rangle ,$ $\left\vert
01\right\rangle ,$ $\left\vert 00\right\rangle \}$ is then given by
\begin{equation}
\rho_{s}\approx\left[
\begin{array}
[c]{cccc}%
0 & 0 & 0 & 0\\
0 & \rho_{22} & \rho_{23} & 0\\
0 & \rho_{23}^{\ast} & \rho_{33} & 0\\
0 & 0 & 0 & \rho_{44}%
\end{array}
\right]  . \label{rho_f}%
\end{equation}
where $\rho_{22}=\rho_{33}=\frac{1}{2}M[c_{3}^{\ast}c_{3}]|_{t\rightarrow
\infty},$ $\rho_{23}=-\frac{1}{2}M[c_{3}^{\ast}c_{3}]|_{t\rightarrow\infty}$
and $\rho_{44}=M[c_{4}^{\ast}c_{4}]|_{t\rightarrow\infty}$.

The analytical results shown above can be verified by our numerical simulation
as illustrated in the figure 4.

\begin{figure}[ptb]
\begin{center}
\includegraphics[
trim=0.000000in 0.000000in 0.000000in -0.460489in,
height=2in,
width=3.2113in
]{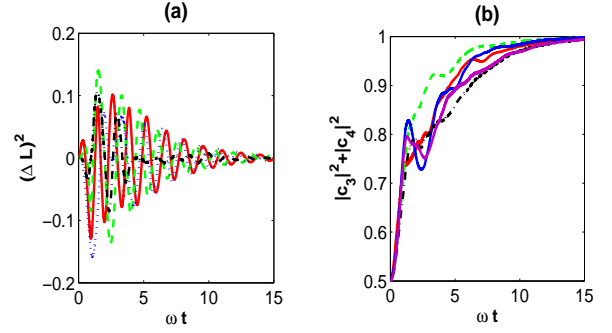}
\end{center}
\caption{Time evolution of the fluctuation $(\Delta L)^{2}$ and the
probability $|c_{3}|^{2}+|c_{4}|^{2}$ for quantum trajectories. Each curve
represents a single quantum trajectory. The parameters are $\omega_{A}%
=\omega_{B}=0.5\omega,$ $\kappa_{A}=\kappa_{B}=1,$ $\gamma=1,J_{xy}=0.5$.}%
\label{single}%
\end{figure}

In Fig. \ref{single} (a), we plot the fluctuation of the system Lindblad
operator $L$ defined as $(\Delta L)^{2}=\left\langle L^{2}\right\rangle
-\left\langle L\right\rangle ^{2}$. It is shown that, as time evolves, the
fluctuation of $L$\ converge to zero in the final steady state for all
trajectories. In Fig. \ref{single} (b), we plot the sum of the probabilities
of $\left\vert \psi_{3}\right\rangle $ and $\left\vert \psi_{4}\right\rangle $
for a few trajectories. The sum of the probability $|c_{3}|^{2}+|c_{4}|^{2}$
will always evolve to $1$ for any single trajectories which means $c_{1}$ and
$c_{2}$ are always zero in the final steady state. The numerical results in
Fig. \ref{single} simply show that the final state of single trajectories take
the form of Eq. (\ref{psi_f}) as predicted by analytical analysis. In
addition, our numerical results show that the final density matrix is
approximately given by
\begin{equation}
\rho_{\mathrm{final}}\approx\left[
\begin{array}
[c]{cccc}%
0 & 0 & 0 & 0\\
0 & 0.25 & -0.25 & 0\\
0 & -0.25 & 0.25 & 0\\
0 & 0 & 0 & 0.5
\end{array}
\right]  . \label{App_rho}%
\end{equation}
which is also consistent with our prediction in Eq. (\ref{rho_f}).

It should be noted that such a steady state has be shown to be the stationary
solution of the Markov master equation \cite{Schneider,Hope}. In our QSD
method, it is showed that this state is also the stationary solution of the
non-Markovian quantum trajectories. Therefore, the ensemble mean of the
quantum trajectories recovers the non-Markovian steady state of the density
matrix. It is important to note that our non-Markovian theory does not depend
on the existence of the non-Markovian master equation.

\begin{figure}[ptb]
\begin{center}
\includegraphics[
trim=0.000000in 0.000000in -0.471890in 0.000000in,
height=2.0081in,
width=3.0113in
]
{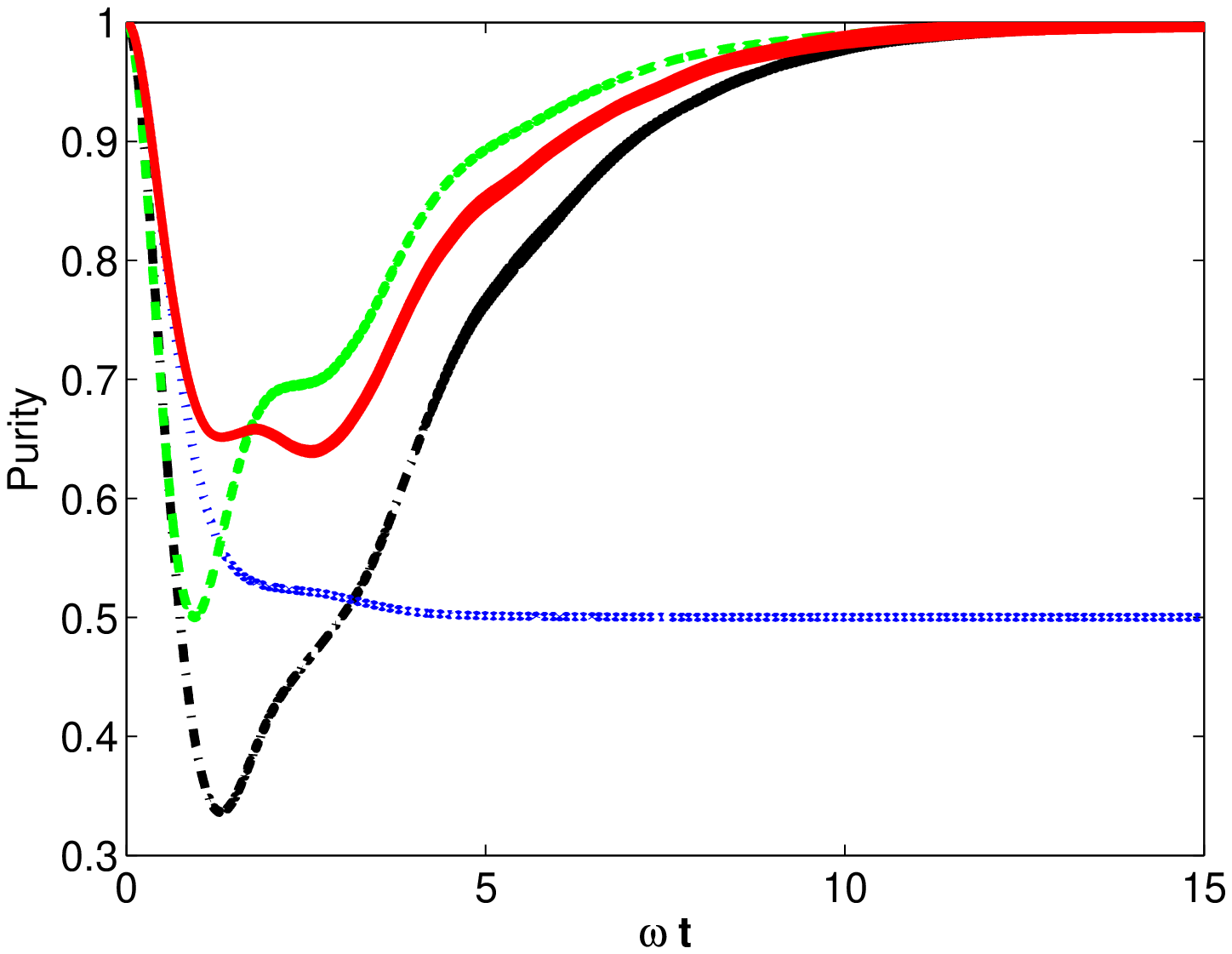}
\end{center}
\caption{Time evolution of purity for different initial states. The initial
states for red (solid), green (dashed), blue (short-dashed) and black
(dash-dotted) lines are $\psi_{0}=\frac{1}{\sqrt{2}}(\left\vert
11\right\rangle +\left\vert 00\right\rangle ),$ $\psi_{0}=\frac{1}{\sqrt{2}%
}(\left\vert 01\right\rangle +\left\vert 10\right\rangle )$, $\psi
_{0}=\left\vert 10\right\rangle ,$ and $\psi_{0}=\left\vert 11\right\rangle $,
respectively. The other parameters are $\omega_{A}=\omega_{B}=0.5\omega,$
$\kappa_{A}=\kappa_{B}=1,$ $\gamma=1,$ $J_{xy}=0.7,$ $J_{z}=0.3.$}%
\label{purity}%
\end{figure}

Finally, we present the time evolution of the purity, defined as
$Purity=tr(\rho^{2})$, which determines whether the quantum state is pure or
mixed. Specifically, we find the purity of the initial state $\psi
_{0}=\left\vert 10\right\rangle ,$ which evolves into a steady state with some
residue entanglement preserved in the final steady state. In Fig.
\ref{purity}, it is validated that the final steady state is a mixed entangled
state since the purity is smaller than one at the steady state. The purity is
approximately $0.5$ in long time limit, which is consistent with Eq.
(\ref{App_rho}). For all the other three initial states, we will see two
qubits will firstly evolve into a mixed (perhaps entangled) state by the
coupling interaction, then decay into the final ground state $\left\vert
00\right\rangle $ (which is a pure state) due to the dissipative environment.

\subsection{Exact versus approximate QSD equations}

\begin{figure}[ptb]
\begin{center}
\includegraphics[
trim=0.000000in 0.000000in 0.000000in -0.807096in,
height=3.0113in,
width=3.0113in
]{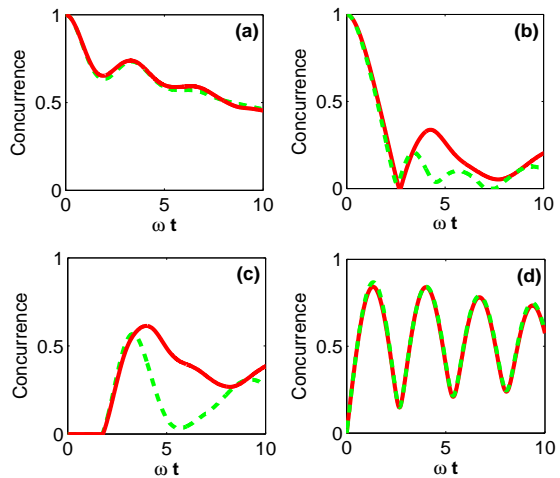}
\end{center}
\caption{Comparison of the dynamics generated by the approximate $O$ operator
and the exact $O$ operator. The red (solid) curve represents the exact QSD
equation, and the green (dashed) curve is for the approximate QSD. (a) (b) (c)
and (d) represent four different initial states $\frac{1}{\sqrt{2}}(\left\vert
10\right\rangle +\left\vert 01\right\rangle )$, $\frac{1}{\sqrt{2}}(\left\vert
11\right\rangle +\left\vert 00\right\rangle )$, $\left\vert 11\right\rangle $
and $\left\vert 10\right\rangle $, respectively. The parameters are chosen as
$\gamma=0.3$, $\omega_{A}=\omega_{B}=0.5\omega,$ $\kappa_{1}=\kappa_{B}=1$,
$J_{xy}=0.5.$ }%
\label{O5}%
\end{figure}

For the coupled two-qubit model, the exact $O$ operator contains five terms,
as shown in Eq. (\ref{Ansztz of O}). It is interesting to investigate the
perturbative QSD equations based on the exact expression of the $O$ operator.
For example, we may drop the first order noise-dependent term $O_{5}$ and only
keep the first four terms, i.e. the $O$ operator in Eq. (\ref{Ansztz of O}) is
approximately written as
\begin{align}
O(t,s,z^{\ast})  &  \approx f_{1}(t,s)O_{1}+f_{2}(t,s)O_{2}\nonumber\\
&  +f_{3}(t,s)O_{3}+f_{4}(t,s)O_{4}.
\end{align}
Clearly, the first order term $O_{5}$ becomes less important when the system
is close to the Markov regimes. Also, numerical calculations will be greatly
simplified if the first order noise is neglected. In Fig. \ref{O5}, we have
compared the exact time evolution of entanglement and the approximate case
without $O_{5}$. The numerical result shows that the accuracy of the zeroth
order approximation is sensitively dependent on the initial state of the
qubits. For the anti-correlated initial states, Fig. \ref{O5} (a) and (d)
shows that the approximation is extremely good compared with the exact result.
On the other hand, for the correlated initial states in case (b) and (c), the
noise dependent term tends to smooth out the entanglement curve. In fact, this
can be explained by analyzing the QSD Eq. (\ref{c1}-\ref{c4}). If the initial
states do not contain $\psi_{1}$ component, i.e. $c_{1}(0)=\left\langle
\psi_{0}|\psi_{1}\right\rangle =0$, then $c_{1}(t)$ is always zero during the
time evolution. The fifth term of the $O$ operator $O_{5}=2\sigma_{-}%
^{A}\sigma_{-}^{B}$ will not affect the dynamics of the system, since the
coefficient of $O_{5}$ , i.e. $\tilde{F}_{5}(t,z^{\ast})$, is always
associated with $c_{1}(t)$. This observation explains our results plotted in
Fig.~{6}. As shown in Fig.~{6}, the noiseless approximation (setting $O_{5}%
=0$) gives extremely good results for the two initial states which do not
contain $\left\vert 11\right\rangle $ component. Our simple analysis here has
already shown the complexity level of the approximate QSD equations arising
from the noise expansion of the $O$ operator. More in-depth study on the
approximate QSD equations will be presented elsewhere.

\section{Conclusion}

In conclusion, the central aim of this paper is to develop a time-local theory
of non-Markovian quantum state diffusion for a model consisting of two
interacting qubits -- a model that is of interest for both quantum information
processing and quantum optics. As an application, we used the non-Markovian
quantum trajectory to analyze the entanglement evolution of the qubit system
for several parameters ranging from strongly non-Markovian to Markov regimes.
Our results also showed explicitly how the environmental memory generates the
entanglement of the qubit system. With the non-Markovian QSD equation, we
further studied the entanglement evolution and the steady entangled state for
the QSD equation. Our time-local QSD approach may be extended to multiple
qubit systems, this will be the topic for a future publication.

\section*{ACKNOWLEDGMENTS}

We would like to thank Professors J. H. Eberly, B. L. Hu and W. T. Strunz for
many interesting conversations and acknowledge support by grants from DARPA
QuEST HR0011-09-1-0008, the NSF PHY-0925174. Part of this work was done while
T.Y. enjoyed the hospitality of the Kavli Institute for Theoretical Physics
China, Beijing.

\end{document}